\documentclass[%
 amsmath,amssymb,
 aps, physrev,
 twocolumn,
 nofootinbib, 
 superscriptaddress, 
 showkeys, 
]{revtex4-2}

\usepackage{graphicx} 
\usepackage{hyperref}
\usepackage{cleveref}
\usepackage{dcolumn}
\usepackage{bm}
\usepackage{color}
\usepackage[normalem]{ulem} 
\usepackage{orcidlink} 

\begin{document}


\title{\boldmath Investigating sub-MeV dark matter annihilation to neutrinos using direct detection experiments}

\author{Boris Betancourt Kamenetskaia\,\orcidlink{0000-0002-2516-5739}}
\email{laybors@ibs.re.kr}
\affiliation{Cosmology, Gravity, and Astroparticle Physics Group, Center for Theoretical Physics of the Universe,
Institute for Basic Science (IBS), Daejeon, 34126, Korea}
\affiliation{Technical University of Munich, TUM School of Natural Sciences, Physics Department, 85748 Garching, Germany}
\affiliation{Max-Planck-Institut f\"ur Physik (Werner-Heisenberg-Institut), Boltzmannstra\ss e 8, 85748 Garching, Germany}

\author{Jong-Chul Park\,\orcidlink{0000-0002-1276-875X}}
\email{jcpark@cnu.ac.kr}
\affiliation{Department of Physics and Institute of Quantum Systems (IQS), Chungnam National University, Daejeon 34134, Republic of Korea}

\author{Merlin Reichard\,\orcidlink{0000-0002-0568-3272}}
\email{m.reichard@tum.de}
\affiliation{Technical University of Munich, TUM School of Natural Sciences, Physics Department, 85748 Garching, Germany}

\author{Gaurav Tomar\,\orcidlink{0000-0002-3468-5306}}
\email{tomar@ktu.edu.in}
\affiliation{School of Basic Sciences and Humanities, APJ Abdul Kalam Technological University, CET Campus, Thiruvananthapuram, Kerala 695016, India}


\begin{abstract}
Dark matter (DM) could self-annihilate into neutrinos in dense regions of the Universe. 
We consider the resulting flux of neutrinos from the Milky Way DM halo and derive exclusion limits on the annihilation cross-section using XENONnT electron recoil data. 
Assuming a $J$-factor independent of the annihilation cross-section, we find leading limits for DM masses below $\mathcal{O}$(MeV). 
Self-annihilating DM
affects the DM halo via dissolution, introducing a cross-section dependency on the halo profile and thus the $J$-factor. 
We discuss such a situation in more detail, finding that the signal rate is below the experimental sensitivity of XENONnT, leaving the annihilation cross-section unconstrained.
\end{abstract}

\maketitle
\flushbottom

\section{Introduction}

The non-gravitational nature of dark matter (DM) is experimentally unknown~\cite{Bertone:2004pz,Bauer:2017qwy,Bertin:2022vfh,Cirelli:2024ssz}, motivating dedicated searches using a variety of experimental strategies, including those targeting products of DM annihilation or decay in dense astrophysical environments of the Universe~\cite{Cirelli:2010xx,Buckley:2013bha,Slatyer:2021qgc}.
DM annihilation into Standard Model (SM) final states such as charged leptons or hadrons is strongly constrained by X-ray and gamma-ray observations from the Milky Way and its satellite galaxies~\cite{Boehm:2002yz,Essig:2013goa, Fermi-LAT:2016uux, Hoof:2018hyn,Cirelli:2023tnx}.
In contrast, annihilation into neutrinos is comparatively less constrained and arises naturally in scotogenic models where DM couples preferentially to the neutrino sector~\cite{Yuksel:2007ac, Farzan:2012sa, Hagedorn:2018spx, Alvey:2019jzx, Blennow:2019fhy, Bell:2020rkw}, or in some supersymmetric extensions of the SM~\cite{Boehm:2013jpa}.

Thermal DM candidates with masses of $\mathcal{O}(\text{MeV})$ are subject to severe cosmological constraints from Big Bang Nucleosynthesis and the Cosmic Microwave Background spectrum~\cite{Boehm:2013jpa, Nollett:2014lwa,Wilkinson:2016gsy,Sabti:2019mhn,Depta:2019lbe,Escudero:2018mvt,Chu:2022xuh,An:2022sva,Chu:2023jyb}, setting a lower limit of $m_\chi > \mathcal{O}(10\,\text{MeV})$. 
However, if DM is produced non-thermally or the early Universe physics differs from the physics describing the DM annihilations in the Milky Way today, the cosmological constraints lose their validity. 
Modifications of the thermal history~\cite{Berlin:2018ztp,Herms:2021fql} or the self-interacting massive particle paradigm~\cite{Hochberg:2014dra,Hochberg:2014kqa} could allow for such a scenario. 
In addition, phase-space considerations of galactic halos~\cite{Tremaine:1979we,Boyarsky:2008ju} set a model-independent lower bound of $\mathcal{O}(\text{keV})$, which, however, can be relaxed if DM consists of more than one degenerate degree of freedom~\cite{Davoudiasl:2020uig}. 
The wash-out of small-scale structures by hot or warm DM results in similar lower bounds~\cite{Viel:2005qj,Boyarsky:2008xj,Viel:2013fqw}. 
Irrespective of the cosmological history of DM, terrestrial constraints offer a direct and complementary handle on DM annihilations today.

Extensive constraints on thermal DM annihilation into neutrinos, particularly in the MeV to PeV range, have been established by numerous neutrino observatories and are comprehensively summarized in Ref.~\cite{Arguelles:2019ouk}. 
However, these constraints predominantly apply to neutrino energies above $\mathcal{O}(\text{MeV})$, leaving the sub-MeV regime largely unconstrained. 
If such sub-MeV neutrinos interact with electrons in terrestrial detectors, they can produce observable signatures in direct detection experiments~\cite{XENON:2020rca, XENON:2022ltv, CDEX:2022mlp, LZ:2023poo}. 
Among these, XENONnT~\cite{XENON:2022ltv} is particularly noteworthy due to its low energy threshold near 1 keV and improved background suppression, which enhances its sensitivity to low-energy new physics. 
With an exposure of 1.16 tonne-years, its recent results mark a significant improvement over previous experiments such as XENON1T~\cite{XENON:2020rca}.

In this work, we establish direct-detection constraints on the DM self-annihilation cross-section into neutrinos for DM masses below $\mathcal{O}(\text{MeV})$, thereby expanding the sensitivity of direct-detection experiments to light dark matter. 
By utilizing the null results from the XENONnT experiment~\cite{XENON:2022ltv} for elastic neutrino-electron scattering (E$\nu$ES), we extend beyond the reach of previous analyses utilizing Borexino~\cite{Arguelles:2019ouk} and 
coherent elastic neutrino-nucleus scattering (CE$\nu$NS) at XENON1T~\cite{Suliga:2024hgp,McKeen:2018pbb}.\footnote{Notably, the first measurements of CE$\nu$NS signals induced by solar $^8$B neutrinos have already been reported by PandaX-4T~\cite{PandaX:2024muv} and XENONnT~\cite{XENON:2024ijk}.}

An essential aspect of this investigation is the role of the astrophysical $J$-factor, which quantifies the line-of-sight integral of the squared DM density in the Milky Way. 
As we will discuss in this article, assuming a $J$-factor independent of the annihilation cross-section (which is denoted as constant from now on) provides robust constraints for DM masses below $\mathcal{O}(\text{MeV})$. However, DM self-annihilations can alter the inner halo structure, reducing the $J$-factor and thereby diminishing the predicted neutrino flux and signal. This phenomenon introduces a dependency of the $J$-factor on the DM annihilation cross-section, necessitating a more detailed examination of its impact on the experimental sensitivity. 
Here, we explore the impact of both a constant $J$-factor approximation and a self-consistent treatment where the $J$-factor varies with the annihilation cross-section, allowing us to quantify the astrophysical uncertainty in our limits.

We also note that significant self-annihilation rates are often accompanied by self-scattering cross-sections, which are known to also change the central DM density. 
This effect, characteristic of self-interacting dark matter (SIDM), can lead to a further dilution of the density profile. 
These profile modifications have been explored in recent studies~\cite{Tulin:2017ara, Yang:2022mxl, Ando:2024kpk, Nadler:2025jwh} and for multi-component DM scenarios~\cite{Kim:2023onk, Kim:2024ltz}. 
In this article, we do not incorporate self-scattering-induced alterations to the DM profile. 
Since such effects would further reduce the central density and thus the signal, our treatment is optimistic in this regard.

The remaining part of this article is organized as follows. 
In Sec.~\ref{sec:neutrino_spectrum}, we discuss the formalism of neutrino-electron scattering, using the neutrino flux from Galactic DM. 
The $J$-factor and its implications are detailed in Sec.~\ref{sec:J_factor}. 
The results are presented in Sec.~\ref{sec:results}, followed by our conclusions in Sec.~\ref{sec:conclusions}.

\section{Neutrino signals from DM annihilations}
\label{sec:neutrino_spectrum}
The signal rate for E$\nu$ES at the experiment can be expressed as  
\begin{equation}
    \frac{d R}{d E_e} = N_T \sum_\ell \int_{E_\nu^\text{min}}^{E_\nu^\text{max}} d E_\nu\, \frac{d\Phi}{d E_\nu} \left( \frac{d\sigma_{\nu_\ell}}{d E_e}  + \frac{d\sigma_{{\bar \nu}_\ell}}{d E_e}\right),
    \label{eq:diffrate}
\end{equation}  
where  
\begin{equation*}
    N_T = Z_{\text{eff}}(E_e) \frac{m_{\text{det}} N_A}{m_{\text{Xe}}}
\end{equation*}  
represents the number of target electrons. 
Here, \( m_{\text{Xe}} \) is the molar mass of xenon, \( m_{\text{det}} \) is the fiducial mass of the detector, and \( N_A \) is the Avogadro's number. 
Additionally, $Z_{\text{eff}}(E_e)$ denotes the number of electrons ionized by an energy deposition $E_e$. 
In synergy with Ref.~\cite{AtzoriCorona:2022jeb}, we approximate this using a series of step functions. 
Alternatively, the relativistic random-phase approximation theory provides a more accurate calculation by incorporating atomic many-body effects~\cite{Chen:2016eab}.

In Eq.~(\ref{eq:diffrate}), $d\Phi/d E_\nu$ is the neutrino flux, and \( d\sigma_{\nu_\ell} / dE_e \) and \( d\sigma_{{\bar \nu}_\ell} / dE_e \) denote the differential cross-sections for neutrino and antineutrino-electron scattering, respectively, given as~\cite{Dev:2021xzd}
\begin{eqnarray}
    \frac{d\sigma_{\nu_{\ell}/{\bar \nu}_{\ell}}}{dE_e} &=& \frac{2G_F^2m_e}{\pi E_\nu^2}\left [c_1^2E_\nu^2+c_2^2(E_\nu-E_e)^2-c_1c_2m_e E_e \right ]\,, \nonumber \\
    && \label{eq:smcont} 
\end{eqnarray}
where \( G_F \), \( E_\nu \), and \( m_e \) denote the Fermi constant, the energy of the incoming neutrino, and the electron mass, respectively. 
The neutrino flavor-dependent coefficients, \( c_1 \) and \( c_2 \), are listed in Table~\ref{tab:coeff}.  
\begin{table}[t]
    \centering
    \begin{tabular}{|c|c|c|}
    \hline
         ~~~Flavor~~~ & ~~~~$c_1$~~~~ & ~~~~$c_2$~~~~  \\
         \hline 
         $\nu_e$ & $s_W^2+\frac{1}{2}$  & $s_W^2$   
         \\
         $\bar{\nu}_e$ & $s_W^2$ & $s_W^2+\frac{1}{2}$  \\
         $\nu_\mu,\nu_\tau$ & $s_W^2-\frac{1}{2}$ & $s_W^2$  \\
         $\bar{\nu}_\mu,\bar{\nu}_\tau$ & $s_W^2$ & $s_W^2-\frac{1}{2}$ \\
         \hline 
    \end{tabular}
    \caption{A summary of the coefficients in Eq.~\eqref{eq:smcont}, where $s_W\equiv \sin\theta_W$ with $\theta_W$ being the Weinberg angle.}
    \label{tab:coeff}
\end{table}

The minimum neutrino energy required to register an electron recoil energy \( E_e \) is given by  
\begin{equation}
  E^{\rm min}_\nu=\frac{E_e+\sqrt{E^2_e+2m_e E_e}}{2}\,,
\end{equation}
while the maximum neutrino energy is set by the mass of the DM particle, \( E_{\nu}^{\rm max} = m_\chi \). 
The efficiency of the XENONnT experiment is taken from Ref.~\cite{XENON:2022ltv}.

For the neutrino flux, we consider DM annihilations in the Milky Way, such that the flux at Earth per neutrino or anti-neutrino per flavor per steradian is given by~\cite{Arguelles:2019ouk}
\begin{equation}\label{eq:diff_flux}
    \frac{d\Phi}{d E_\nu} = \frac{1}{2}\frac{1}{4\pi} \frac{\langle \sigma v \rangle}{\kappa m_\chi^2} \frac{1}{3} \frac{dN_\nu}{dE_\nu} J(\Omega)\,,
\end{equation}
where $\kappa = 2$ ($\kappa = 4$) for a Majorana (Dirac) DM candidate $\chi$, annihilating into a pair of neutrino and anti-neutrino with thermally averaged cross-section $\langle \sigma v \rangle$, and $J(\Omega)$ is the $J$-factor, corresponding to the line-of-sight integral between the Earth and the source.
The neutrino spectrum at the source reads
\begin{equation}
    \frac{d N_\nu}{d E_\nu} = \frac{2 m_\chi}{E_\nu^2} \delta(1-E_\nu/m_\chi)\,,
\end{equation}
for which we assume a flavor composition of $(\nu_e:\nu_\mu:\nu_\tau)=(1:1:1)$. Neutrinos undergo flavor oscillations during propagation from outer regions of the Milky Way to the Earth. Given the large distance ($\sim \rm kpc$) and the considered sub-MeV-scale energies of neutrinos in our scenario which lead to an oscillation length of $L_{\rm osc}\sim10^4~\rm km$, vacuum oscillations fully average out. Hence, the final flux keeps the same composition regardless of the initial production mechanism~\cite{Arguelles:2019ouk}. 
Furthermore, we neglect Earth matter effects, as the Mikheyev-Smirnov-Wolfenstein (MSW) resonance energy in Earth’s mantle ($\sim$ GeV) far exceeds the sub-MeV energy range of interest in our scenario. At these energies, the matter-induced potential is subdominant to vacuum oscillation terms~\cite{Wolfenstein:1977ue,Mikheev:1986wj,Bahcall:2004mz}.

Extragalactic sources also contribute to the neutrino flux at the Earth~\cite{Beacom:2006tt,Moline:2014xua}; however, we find this contribution to be negligible, as these neutrinos are redshifted below the experimental energy threshold.

A key component in determining the expected neutrino flux is the Galactic $J$-factor, which we discuss in the next section.

\section{Galactic $J$-factor}
\label{sec:J_factor}
A variety of models have been proposed to describe the DM halo density distribution in galaxies, motivated by both cosmological simulations and dynamical observations. Common examples include cuspy profiles such as the Navarro-Frenk-White (NFW)~\cite{Navarro:1995iw,Navarro:1996gj} and Einasto~\cite{Graham:2005xx,Navarro:2008kc} profiles, as well as cored distributions that may arise from baryonic feedback or self-interactions, such as the isothermal~\cite{Begeman:1991iy} or Burkert~\cite{Burkert:1995yz} profiles. In our analysis, we allow for any of the aforementioned spherically symmetric DM density profiles, denoted $\rho_i(r)$.

\begin{figure}[t!]
    \centering
\includegraphics[width=0.48\textwidth]{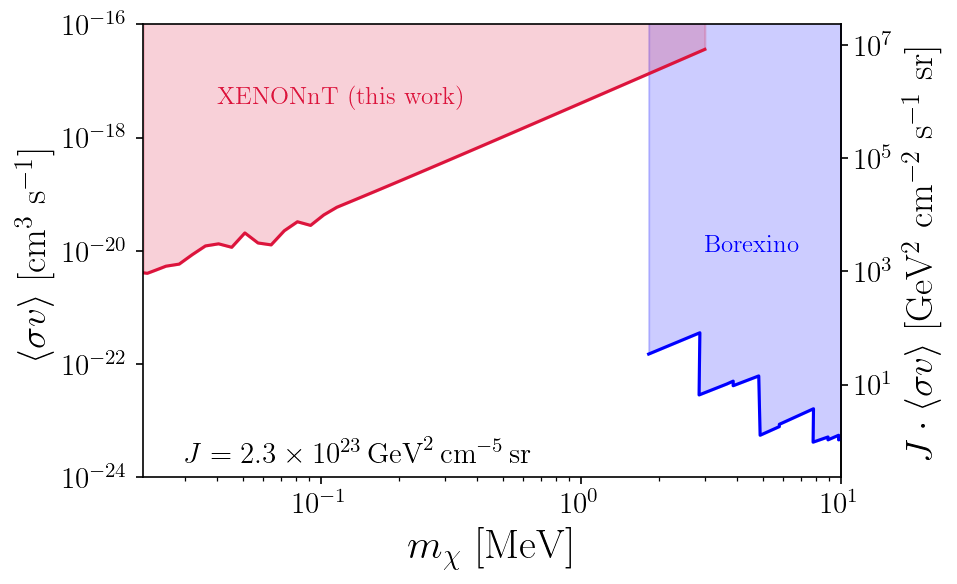}
	\caption{The $90\%$ C.L. exclusion limit on the thermally averaged Majorana DM annihilation cross-section into neutrinos from XENONnT for $J=2.3\times 10^{23}$ GeV$^2$cm$^{-5}$sr. 
    The blue-shaded region represents the exclusion limit from the Borexino experiment~\cite{Arguelles:2019ouk}.}
	\label{fig:exclusion_limit_sigmav}
\end{figure}

For non-zero DM annihilation cross-sections, the density profile of Galactic DM is expected to decrease as the Galaxy evolves in time due to the continuous depletion of DM particles. 
This process is known as dissolution. 
For such an annihilation process with two DM particles in the initial state, the change in the density $\rho_\chi(r,t)$ is described by the master equation~\citep{Ahn:2007ty}
\begin{equation}
    \frac{1}{m_\chi}\frac{\partial \rho_\chi(r,t)}{\partial t}=-\langle\sigma v\rangle \left(\frac{\rho_\chi(r,t)}{m_\chi}\right)^2.
\end{equation}
Taking the DM profile $\rho_i$ as the initial condition, the current DM density distribution is given by
\begin{align}\label{eq:dens_prof}
    \rho_\chi(r)&=\frac{\rho_{i}(r)}{[1+\left(\rho_{i}(r)/\rho_{\mathrm{sat}}\right)]}\,,
\end{align}
where
\begin{align}\label{eq:sat_dens}\small
    &\rho_{\mathrm{sat}}=\frac{m_\chi}{\langle\sigma v\rangle t_{\rm age}}\cr
    &\simeq3.17~\frac{\mathrm{MeV}}{\mathrm{cm}^3}\left(\frac{m_\chi}{1~\rm MeV}\right)\left(\frac{\langle \sigma v\rangle}{10^{-18}~\frac{\mathrm{cm}^3}{\rm s}}\right)^{-1}\left(\frac{t_{\rm age}}{10~\rm Gyr}\right)^{-1}~~
\end{align}
can be understood as a characteristic density at which annihilations become significant. 
From Eq.~\eqref{eq:dens_prof}, we see that when $\rho_{i}\gg\rho_{\rm sat}$, the DM density $\rho_\chi$ saturates, leading to the formation of a core-like structure near the Galactic Center (GC). The age of the Milky Way is denoted by $t_{\rm age}$, taken as 10 Gyr in our analysis.

We note that for the annihilation cross-sections of interest in this work, the dissolution effect washes out the differences between various DM profiles. Once the central density exceeds $\rho_{\rm sat}$, all profiles tend toward a similar cored structure, making our results largely independent of the initial choice of the halo model (e.g., NFW vs. Burkert).

The presence of the supermassive black hole (SMBH) Sgr A$^*$ at the center of the Milky Way, with an estimated mass of $M_{\rm BH}  \simeq  4.15  \times  10^6M_\odot$~\cite{Gravity:2019nxk}, has been considered in theoretical studies. 
The growth of the SMBH could, in principle, lead to an enhancement of the DM density profile, forming a so-called \textit{DM spike}~\cite{Gondolo:1999ef}. 
However, for the range of annihilation cross-sections considered in this study, DM annihilations fully suppress any such enhancement, making the resulting density distribution indistinguishable from the initial density profile with dissolution effects.

In order to summarize the astrophysical effects related to the DM density profile, it is customary to introduce the $J$-factor (in units of GeV$^2\,$cm$^{-5}\,$sr)
\begin{align}
  J=\int{d\Omega \int_{\rm  l.o.s.}  d  s\,  
  \rho_\chi  (r ( s,  \theta ) )^2}\,,
\end{align}
where $s$ is the line of sight (l.o.s.) coordinate, which is the distance measured from local Earth coordinates. Under this coordinate transformation from the GC to the Earth, the radial coordinate is expressed as $r( s, \theta) = \sqrt{r_\odot^2  + s^2 - 2 r_\odot s \cos \theta}$, where $\theta$ is the angle between the l.o.s. direction and the Earth-GC axis, and $r_\odot\sim8.33~\rm kpc$ is the distance between the Sun and the GC. We note an explicit dependence of the $J$-factor on the solid angle; however, since DM direct-detection experiments typically lack directional sensitivity, we compute this quantity via an all-sky angular integration.
For the Galactic $J$-factor, we consider two scenarios:
\begin{enumerate}
    \item[i)] A constant $J$-factor: arising from the assumption of no DM dissolution that is independent of the annihilation cross-section $\langle\sigma v\rangle$; this corresponds to the standard literature approach where the DM density profile remains unaltered by annihilation effects. We use the all-sky $J$-factor $J = 2.3\times 10^{23} \,\text{GeV}^2\,\text{cm}^{-5}\,\text{sr}$, as obtained in Ref.~\cite{Arguelles:2019ouk}.
    \item[ii)] A depletion-dependent $J$-factor: resulting from the inclusion of the effect of DM depletion due to the annihilation, hence explicitly depending on $\langle\sigma v\rangle$, as summarized in this section.
\end{enumerate}
In the following section, we present our results based on a $\chi^2$ fit.

\section{Results}
\label{sec:results}

\begin{figure}[t!]
\centering
\includegraphics[width=0.48\textwidth]{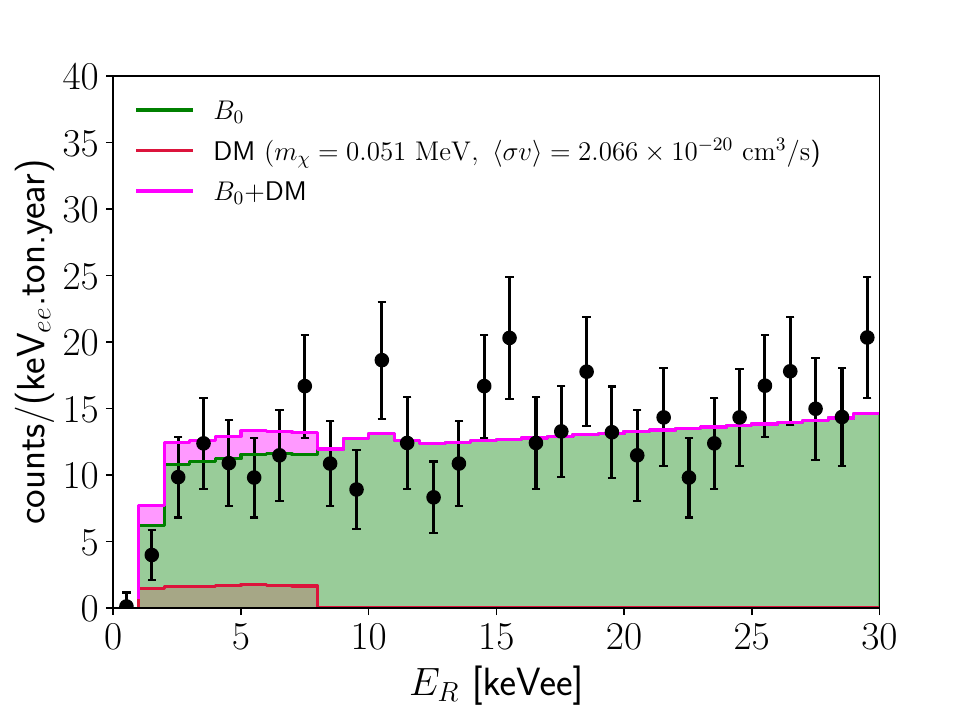}
\caption{Expected DM-neutrino annihilation signal compared to XENONnT data~\cite{XENON:2022ltv}. 
Black dots with error bars represent XENONnT data, while $B_0$ denotes the background from Ref.~\cite{XENON:2022ltv}.}
\label{fig:recoil_spectrum}
\end{figure}

\begin{figure}[t!]
\centering
\includegraphics[width=0.48\textwidth]{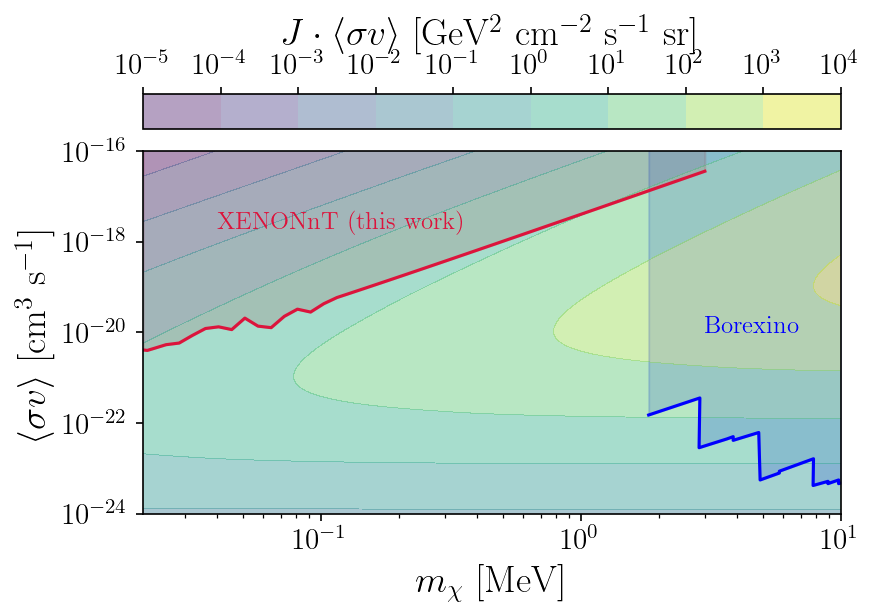}
\caption{The contours show $J\cdot \langle \sigma v\rangle$ as a function of $m_\chi$ and $\langle \sigma v\rangle$ for $\kappa = 2$ and a Milky Way profile including the dissolution effect.}
	\label{fig:dissolution}
\end{figure}

We constrain the DM-neutrino annihilation cross-section by employing a Gaussian $\chi^2$ function 
\begin{equation}
 \chi^2(\mathcal{S})=\sum_{k}\left(\frac{R^k_{\rm exp}-R^k_{{\rm DM}+B_0}(\mathcal{S})}{\sigma_k}\right)^2\,.
 \label{eq:chi2_pen}
\end{equation}
Here, \( k \) runs over the energy bins, and \( \sigma_k \) denotes the statistical uncertainty per bin for the given experiment. 
We consider 30 bins for our XENONnT analysis, following the published result~\cite{XENON:2022ltv}.

In Eq.~\eqref{eq:chi2_pen},
$R^k_{\rm exp}$ represents the observed differential event rate for the XENONnT experiment, while
\begin{equation*}
R^k_{{\rm DM}+B_0} (\mathcal{S})=R^k_{{\rm DM}} + B^k_0\,,
\end{equation*}
where $B^k_0$ denotes the background reported by the XENONnT collaboration~\cite{XENON:2022ltv} and whose components are summarized in Tab.~I in Ref.~\cite{XENON:2022ltv}. Neutron‐induced nuclear recoils are effectively eliminated by the electron-recoil/nuclear-recoil discrimination of the experiment, which exceeds $99\%$ and therefore do not enter in $B_0^k$. Potential contributions from atmospheric neutrinos and the diffuse supernova neutrino background are negligible, as both populations produce recoil spectra peaked above the $1–30~\rm keV$ region of interest (at $\sim$ GeV and $\sim10$ MeV, respectively), yielding event rates orders of magnitude below the solar neutrino electron-recoil component.

Finally, $R^k_{{\rm DM}}$ describes the additional signal due to annihilating DM in the center of the Galaxy, calculated using Eq.~(\ref{eq:diffrate}) for each energy bin.
The bound on the DM-neutrino annihilation cross-section is obtained by solving
\begin{equation}
 \chi^2-\chi^2_{\rm min}=2.71\,,
 \label{eq:chi2_min}
\end{equation}
which, for one degree of freedom, corresponds to  90$\%$ C.L. sensitivity.

In the naive scenario i) of Sec.~\ref{sec:J_factor}, the $J$-factor is unperturbed by the DM annihilations, such that we would expect the quantity $J\cdot\langle \sigma v\rangle$ to rise linearly with increasing values of $\langle\sigma v\rangle$. 
Using the outlined methodology, we find for the all-sky $J$-factor of $J = 2.3\times 10^{23} \,\text{GeV}^2\,\text{cm}^{-5}\,\text{sr}$ the exclusion limit on the thermally averaged Majorana DM annihilation cross-section into neutrinos depicted in Fig.~\ref{fig:exclusion_limit_sigmav}. We note that the limits would worsen by a factor of 2 in the case of Dirac DM, as inferred from the differential neutrino flux 
of Eq.~\eqref{eq:diff_flux}.

We present in Fig.~\ref{fig:recoil_spectrum} the recoil spectrum, including the experimental data from XENONnT~\cite{XENON:2022ltv}, the background, and the potential DM-induced signal we consider in this work, corresponding to the best-fit value of the $\chi^2$-minimization. Finally, we would like to point out that the choice of presenting limits in terms 
of the product $J\cdot\langle\sigma v\rangle$ is particularly convenient, since 
all of the halo model dependence, and hence the full astrophysical uncertainties,
is encapsulated in the $J$-factor. Our results can be readily applied to any particular Milky Way profile by simply rescaling our curves by the ratio of the particular $J$ to $2.3\times10^{23}\,\mathrm{GeV}^2\mathrm{cm}^{-5}\mathrm{sr}$, without redoing the full analysis.

In the more realistic scenario ii), however, the dissolution effect affects the DM density distribution, as captured in Eq.~\eqref{eq:dens_prof}, and subsequently, the $J$-factor. 
As $\langle\sigma v\rangle$ increases, the local DM density in the central region deviates from the initial prediction due to annihilation-driven dissolution, which becomes significant for $\rho_{i}(r)\gtrsim\rho_{\rm sat}$. 
As the cross-section grows, the saturation density decreases (see Eq.~\eqref{eq:sat_dens}), thereby lowering the threshold at which the central density becomes significantly depleted. 
This results in an increasingly pronounced flattening of the density profile in the inner Galaxy, where annihilation rates are highest.

We show in Fig.~\ref{fig:dissolution} the predicted values of $J\cdot \langle \sigma v\rangle$ as a function of $m_\chi$ and $\langle \sigma v\rangle$ for Majorana DM, highlighting the non-trivial dependency that the dissolution effect introduces.
For fixed $m_\chi$, this product exhibits a peak beyond which increasing the annihilation cross-section leads to a decline. 
This turnover arises because enhanced annihilation depletes the central DM density, suppressing the overall signal. 
At high cross-sections, the number density becomes sufficiently low that further increases in $\langle \sigma v \rangle$ no longer boost the flux, as the $J$-factor flattens and eventually decreases. 
Indeed, we note that if the dissolution effect is included, the neutrino flux at Earth is below the experimental limits derived in this work by at least $\sim4$ orders of magnitude, resulting in an inaccessible prediction with current detector technologies.
Our approach, which neglects self-scattering effects and thus adopts an optimistic estimate of the $J$-factor, already yields fluxes that fall well below current experimental sensitivity. 
Including the additional suppression from core formation would only strengthen this conclusion.
However, the Borexino limits reported in Ref.~\cite{Arguelles:2019ouk} can probe even such a scenario for $m_\chi\gtrsim 1\,\text{MeV}$, as indicated by the contours in Fig.~\ref{fig:dissolution}.

\section{Conclusions}
\label{sec:conclusions}
In this article, we calculated the exclusion limit on the DM-neutrino annihilation cross-section $\langle \sigma v\rangle$ using the absence of an abnormal flux of neutrino-induced electron scattering events at XENONnT. 
We find that for the standard approach of using a constant $J$-factor, XENONnT constraints for $m_\chi \simeq 20\,\text{keV}$ cross-sections $\langle \sigma v\rangle\gtrsim 4.0\times 10^{-21}\,\text{cm}^3\,\text{s}^{-1}$, reaching up to $\langle \sigma v\rangle\gtrsim 3.6\times 10^{-17}\,\text{cm}^3\,\text{s}^{-1}$ for $m_\chi = 3\,\text{MeV}$. 
For a non-constant $J$-factor, the inclusion of the dissolution effect in the halo profile implies a significant reduction of the predicted flux at the Earth, resulting in the inability to probe the annihilation cross-sections at XENONnT for the DM masses considered in this work.  
Therefore, the current experimental sensitivity must increase significantly to probe sub-MeV scale DM. 
Meanwhile, for DM above the MeV scale, we find that current generation experiments, such as Borexino, can already probe non-standard Galactic halo profiles affected by the dissolution effect.

\section*{Acknowledgments}
The authors would like to thank the organizers of PPC 2024 at IIT Hyderabad, where the idea gained its initial momentum.
The work of BBK and MR was supported by the Deutsche Forschungsgemeinschaft (DFG, German Research Foundation) under Germany’s Excellence Strategy - EXC-2094 - 390783311 and under the Collaborative Research Center SFB1258 grant - SFB-1258 - 283604770. BBK is supported by IBS under the project code IBS-R018-D3. The work of JCP is supported by the National Research Foundation of Korea grant funded by the Korea government (MSIT) [RS-2024-00356960].

%

\end{document}